\documentclass{rspublic}
\usepackage{amssymb,amsmath}
\usepackage{multirow}
\usepackage{graphicx}
\usepackage{array}
\usepackage{rotating,booktabs}
\usepackage{lscape}

\begin{document}

\title[]{Interconnections between various analytic approaches applicable to third-order nonlinear differential equations}

\author[Mohanasubha, Chandrasekar, Senthilvelan and Lakshmanan]{R. Mohanasubha$^1$, V. K. Chandrasekar$^2$, M. Senthilvelan$^1$ and \\M. Lakshmanan$^1$}

\affiliation{$^1$ Centre for Nonlinear Dynamics, School of Physics,
Bharathidasan University, Tiruchirappalli - 620 024, India\\
$^2$ Centre for Nonlinear Science and Engineering, School of Electrical and Electronics Engineering, SASTRA University, Thanjavur - 613 401, India}

\label{firstpage}

\maketitle

\begin{abstract}{Third-order nonlinear differential equations, Null forms, Symmetries, Darboux polynomials, Integrating factors and Jacobi last multipliers} 
We unearth the interconnection between various analytical methods which are widely used in the current literature to identify integrable nonlinear dynamical systems described by third-order nonlinear ordinary differentiable equations (ODEs). We establish an important interconnection between extended Prelle-Singer procedure and $\lambda$-symmetries approach applicable to third-order ODEs to bring out the various linkages associated with these different techniques. By establishing this interconnection we demonstrate that given any one of the quantities as a starting point in the family consisting of Jacobi last multipliers, Darboux polynomials, Lie point symmetries, adjoint-symmetries, $\lambda$-symmetries, integrating factors and null forms one can derive the rest of the quantities in this family in a straightforward and unambiguous manner. We also illustrate our findings with three specific examples.

\end{abstract}

\section{Introduction}
In a previous paper (Mohanasubha \textit{et al.} Proc. R. Soc. A 2014) we have established a connection between extended Prelle-Singer procedure with five other analytical methods which are widely used to identify integrable systems described by second-order ODEs and brought out the interconnections between Lie point symmetries, $\lambda$-symmetries, Darboux polynomials, Prelle-Singer procedure, adjoint-symmetries and Jacobi last multiplier. We have also illustrated the interconnections by considering a nonlinear oscillator equation as an example. A natural question which arose was whether these interconnections exist in higher order ODEs as well. In this paper we consider this problem for third-order ODEs and come out with certain new and interesting results.

We start our investigation with the following question. Given any one of the quantities as a starting point in the family, consisting of multipliers, Darboux polynomials, Lie  point symmetries, adjoint-symmetries, $\lambda$-symmetries, integrating factors and null forms can one obtain the rest of all quantities in this family? In the case of third-order ODEs exploring the interconnections between the above said quantities is difficult both conceptually as well as technically which will be clear as we proceed.  In fact, we get an affirmative answer to our questions. We divide our analysis into two parts. In the first part, as we did in the case of second-order ODEs we try to interlink the quantities by introducing suitable transformations in the extended Prelle-Singer procedure. Once the interconnections are identified, in the second part, we demonstrate that starting from any one of the quantities in this family one can derive all the other quantities in a simple and unambiguous way.

In the first part, the interplay between various methods is brought out by establishing a road map between extended the Prelle-Singer procedure and the other methods, namely (i) Jacobi last multipliers, (ii) Darboux polynomials, (iii) Lie point symmetries, (iv) adjoint-symmetries and (v) $\lambda$-symmetries. In the above we have listed out the methods in  chronological order. In other words JLM was the one invented first and the last one in this direction is the extended Prelle-Singer procedure. So naturally we try to accommodate other methods into extended Prelle-Singer procedure. In the later procedure one has three equations to integrate to obtain the null forms $U$ and $S$ and the integrating factor $R$. To interconnect these quantities we introduce three new variables $V, X$ and $F$, which are to be determined through the transformations, $U=-\frac{D[V]}{V},~~S=\frac{X}{V}$ and $R=\frac{V}{F}$ so that one can rewrite the determining equations for $U,S$ and $R$ in terms of $X, V$ and $F$, respectively. Doing so, we find that the determining equation for the function $F$ exactly coincides with the determining equation for the Darboux polynomials which in turn establishes a connection between the integrating factors and Darboux polynomials. Now properly combining the expressions which arise in the extended Prelle-Singer procedure with the $\lambda$-symmetry determining equation we obtain the following relation between $U$ and $S$ with $\lambda$ (see Sec.3 for details), namely $U=-\frac{(D[\lambda]+\lambda^2+S)} {\lambda}$. This interlink is vital and new to the literature. Since the null forms $U$ and $S$ are now expressed in terms of $V$ and $X$, this interlink can also be written in terms of $V$ and $\lambda$. Doing so we obtain an equation solely in terms of $V$ and $\lambda$, that is $D^2[V]-D[V](\phi_{\ddot{x}}+\lambda)+V(D[\lambda]+\lambda^2-\phi_{\dot{x}})=0$ which in turn determines $\lambda$ if $V$ is known or vice versa (Here $\phi(t,x,\dot{x},\ddot{x})$ is the function defining the third-order ODE, see Eq.(\ref{main1}) below). We assume that $\lambda$ is known. In this case $V$ can be determined by solving this equation and from the latter $X$ can be fixed and so $U,S$ and $R$ can be determined. The Darboux polynomials can be determined either from the multipliers or from their own determining equations. This in turn establishes the interconnection between all these methods. One important result which we observe in this part is that for a given $\lambda$, the null forms $U$ and $S$ are not unique which can be seen from the above relation. Because of this, one may find multiple null forms for a given equation and one has to choose the correct form of $U$ and $S$ in order to obtain an independent integral. From the remaining forms of $U$ and $S$ one may obtain other useful information such as symmetries, integrating factors, Darboux polynomials and Jacobi last multiplier; however the resulting integrals turn out to be dependent ones. This feature is different from second-order ODEs where one has one-to-one correspondence between the null forms and $\lambda$-symmetries ($\lambda=-S$). We illustrate this important point by considering three examples.

Next we move on to the question that given any one of the system quantities how to derive the remaining quantities in this family. Using the interconnections which we identified above, and starting from any one of the quantities, we establish road maps to connect to all the other quantities. After a detailed analysis we find that it is sufficient to consider any one of the following three cases as starters, namely (i) Lie point symmetries, (ii) integrating factors or (iii) Darboux polynomials. Then the remaining quantities can be uniquely determined. The other three possibilities, namely (iv) $\lambda$-symmetries, (v) null forms and (vi) JLM are only subcases of the previous three cases, respectively. The exact road map in all the cases is explicitly demonstrated in Sec.4. To illustrate the ideas introduced in this paper we consider three examples. 

 Since some of the methods applicable for the third-order ODEs are not much discussed in the literature as in the case of second-order ODEs, to begin with we review briefly the six different analytical methods in the next section (Sec.2) in order to be self contained.

\section{Methods}
In this section we briefly recall the six analytical methods with reference to the third-order ODEs which are widely used in the contemporary literature to derive one or more of the following quantities, namely (i) multipliers, (ii) integrating factors, (iii) symmetries and (iv) integrals. We present the methods in chronological order.

\subsection{Jacobi last multiplier method}
The Jacobi last multiplier method was introduced by Jacobi in the year 1844 (Jacobi, 1844,1886), but laid dormant for a long time. Nucci \textit{et al.} have demonstrated the applicability of this method in exploring non-standard Lagrangians associated with certain second-order nonlinear ODEs (Nucci \& Leach, 2008). The attractive feature of the method is that if we know two independent Jacobi last multipliers $M_1$ and $M_2$ of the given equation, then their ratio yields a first integral for the given equation. 

We consider a third-order ODE, 
\begin{equation}
\dddot{x}=\phi(t,x,\dot{x},\ddot{x}),\label{main1}
\end{equation}
where $\phi$ is a function of $t,x,\dot{x}$ and $\ddot{x}$. We rewrite the third-order ODE (\ref{main1}) equivalently as a system of following three first-order ODEs in the absence of explicit $t$-dependence
\begin{equation}
\dot{x}=f(x,y,z),~~\dot{y}=g(x,y,z),~~\dot{z}=h(x,y,z),\label{jlmeqns}
\end{equation}
where $f,~g$ and $h$ are suitably chosen functions of $x,~y$ and $z$.  The analysis can be extended in principle to the general case with $t$-dependence included, though it becomes more involved now (see for example, Clebsch, 2009). We consider two integrals $I_1(x,y,z)$ and $I_2(x,y,z)$ of (\ref{jlmeqns}) whose total differentials are given by 
\begin{equation}
\dot{x}\frac{\partial I_1} {\partial x}+\dot{y}\frac{\partial I_1} {\partial y}+\dot{z}\frac{\partial I_1} {\partial z}=0,~~\dot{x}\frac{\partial I_2} {\partial x}+\dot{y}\frac{\partial I_2} {\partial y}+\dot{z}\frac{\partial I_2} {\partial z}=0. 
\end{equation}
The equations of motion which satisfy these two conditions can be written as
\begin{eqnarray}
\dot{x}=\bigg(\frac{\partial I_1} {\partial y}\frac{\partial I_2} {\partial z}-\frac{\partial I_1} {\partial z}\frac{\partial I_2}{\partial y} \bigg)=f_1(x,y,z),\nonumber \\
\dot{y}=\bigg(\frac{\partial I_1} {\partial z}\frac{\partial I_2} {\partial x}-\frac{\partial I_1} {\partial x}\frac{\partial I_2}{\partial z} \bigg)=f_2(x,y,z),\nonumber \\
\dot{z}=\bigg(\frac{\partial I_1} {\partial x}\frac{\partial I_2} {\partial y}-\frac{\partial I_1} {\partial y}\frac{\partial I_2}{\partial x} \bigg)=f_3(x,y,z),\label{jlmeqns1}
\end{eqnarray}
where we have defined for simplicity the expressions inside the parenthesis as $f_1(x,y,z),~f_2(x,y,z)$ and $f_3(x,y,z)$. We assume that these equations be the ones coming out from the original equation (\ref{jlmeqns}) after multiplying by an integrating factor $M(x,y,z)$. Then comparing Eqs.(\ref{jlmeqns}) and (\ref{jlmeqns1}), we have
\begin{equation}
f_1=Mf(x,y,z),~~f_2=Mg(x,y,z),~~f_3=Mh(x,y,z).\label{comp}
\end{equation}
Interestingly one may also observe from (\ref{jlmeqns1}) that
\begin{equation}
\frac{\partial f_1} {\partial x}+\frac{\partial f_2} {\partial y}+\frac{\partial f_3} {\partial z}=0.\label{condi1}
\end{equation}
Substituting (\ref{comp}) in (\ref{condi1}) and expanding it we can get the following equation for the multiplier $M$, namely
\begin{equation}
f(x,y,z)\frac{\partial M} {\partial x}+g(x,y,z)\frac{\partial M} {\partial y}+h(x,y,z)\frac{\partial M} {\partial z}+\bigg(\frac{\partial f} {\partial x}+\frac{\partial g} {\partial y}+\frac{\partial h} {\partial z} \bigg)M=0.\label{mmmm}
\end{equation}
Eq.(\ref{mmmm}) can be simplified to yield 
\begin{equation}  
\hat{D}[\log M]+\bigg(\frac{\partial f} {\partial x}+\frac{\partial g} {\partial y}+\frac{\partial h} {\partial z} \bigg)=0,\label{meet20}
\end{equation}
where the differential operator $\hat{D}=\dot{x}\frac{\partial}{\partial x}+\dot{y}\frac{\partial}{\partial y}+\dot{z} \frac{\partial}{\partial z}$.
Choosing $f(x,y,z)=y$, $g(x,y,z)=z$ and $h(x,y,z)=\phi$, Eq.(\ref{meet20}) can be written as
\begin{equation}  
\hat{D}[\log M]+\phi_{\ddot{x}}=0.\label{met20}
\end{equation}
Substituting the given equation in (\ref{met20}) and solving the resultant equation one can obtain the multiplier associated with a given third-order ODE (Jacobi, 1844,1886; Nucci \& Leach, 2008). Note that there can be more than one multiplier for a given ODE (\ref{main1}). We further remark here  that when the function $\phi$ in (\ref{main1}) also depends on $t$ explicitly, the above operator $\hat{D}$ will be replaced by the total differential operator $D=\frac{\partial}{\partial t}+\dot{x}\frac{\partial}{\partial x}+\dot{y}\frac{\partial}{\partial y}+\dot{z} \frac{\partial}{\partial z}$.

An important application of multipliers is that one can determine the integrals associated with the given equation by evaluating their ratios. If $M_1$ and $M_2$ are two multipliers, then it is straightforward to check from (\ref{met20}) that one can identify an integral as $I=\frac{M_1}{M_2}$ (Nucci, 2005). So if we have sufficient number of multipliers we can obtain the necessary integrals to prove the integrability of (\ref{main1}).

\subsection{Darboux polynomials approach}
Darboux polynomials method was developed by Darboux in the year 1878 (Darboux, 1878). It provides a strategy to find first integrals. Darboux showed that if we have $\frac{n(n+1)} {2}+2$ Darboux polynomials, where $n$ is the order of the given equation, then there exists a rational first integral which can be expressed in terms of these polynomials. In the case of third-order ODEs, we can get eight Darboux polynomials for a given equation (Darboux, 1878). 

The Darboux polynomials determining equation is given by the following expression
\begin{equation}
 D[F]=g(t,x,\dot{x},\ddot{x})F,\label{Darb}
\end{equation}
where $D$ is the total differential operator and $g(t,x,\dot{x},\ddot{x})$ is the cofactor (Darboux, 1878). Note also that for a given $F$, and an integral $I$ of (\ref{main1}), the quantity $f(I) F$, where $f$ is arbitrary, is also a solution of (\ref{Darb}) for the same cofactor. Using this fact and solving Eq.(\ref{Darb}), we can obtain the Darboux polynomials $(F)$ and the cofactors $(g)$, see Mohanasubha \textit{et al.} (2014a) for further details on the method. 
                      
\subsection{Lie symmetry analysis}
Lie symmetry analysis is one of the powerful methods to investigate the integrability property of the given ODE of any order $n$ where one first explores the symmetry vector fields associated with it. The Lie symmetry vector fields can then be used to derive integrating factors, conserved quantities and so on (Olver 1993).

We consider a third-order ODE of the form (\ref{main1}). The invariance of Eq.(\ref{main1}) under an one parameter group of Lie point symmetries, corresponding to infinitesimal transformations, 
\begin{equation}
T=t+\varepsilon \,\xi(t,x),~~~X=x+\varepsilon \,\eta(t,x),\quad \epsilon \ll 1,\label{asm}
\end{equation}
where $\xi(t,x)$ and $\eta(t,x)$ are the coefficient functions of the associated generator of an infinitesimal transformation and $\varepsilon$ is a small parameter, is specified by
\begin{equation}
\xi \frac{\partial \phi}{\partial t}+\eta \frac{\partial \phi}{\partial x}+\eta^{(1)}\frac{\partial \phi}{ \partial \dot x}+\eta^{(2)}\frac{\partial \phi}{ \partial \ddot x}-\eta^{(3)}=0.\label{liec}
\end{equation} 
Here $\eta^{(1)},~\eta^{(2)}$ and $\eta^{(3)}$ are the first, second and third prolongations, respectively, of the infinitesimal point transformations (\ref{asm}) and are defined to be  
\begin{equation}
\eta^{(1)}=\dot{\eta}-\dot{x}\dot{\xi},~~ \eta^{(2)}=\dot{\eta}^{(1)}-\ddot{x}\dot{\xi},~~\eta^{(3)}=\dot{\eta}^{(2)}-\dddot{x}\dot{\xi}.\label{prol}
\end{equation}
In the above over dot denotes total differentiation with respect to $t$.
Substituting the known expression $\phi$ in (\ref{liec}) and solving the resultant equation we can get the Lie point symmetries associated with the given third-order ODE. The maximum number of admissible Lie point symmetries for a third-order ODE (\ref{main1}) is seven (Olver 1993). The associated vector field is given by $\Omega=\xi \frac{\partial} {\partial t}+\eta\frac{\partial} {\partial x}$. 

One may also introduce a characteristics $Q=\eta-\dot{x}\xi$ and rewrite the invariance condition (\ref{liec}) in terms of a single variable $Q$ in the form
\begin{equation}  
D^3[Q]  = \phi_{\ddot{x}} D^2[Q]+\phi_{\dot{x}} D[Q]+\phi_x Q.
\label{met16}
\end{equation}
One can deduce the coefficient functions $\xi$ and $\eta$ associated with the Lie point symmetries from out of the class of solutions to (\ref{met16}) which depends only on $x$ and $t$, and also has a linear dependence in $\dot{x}$.

\subsection{Adjoint-symmetries}
In general, for systems of one or more ODEs, an integrating factor is a set of functions, multiplying each of the ODEs, which yields a first integral. If the system is self-adjoint (that is the adjoint of the linearized symmetry condition is the same as the linearized symmetry condition) then its integrating factors are necessarily solutions of its linearized system. Such solutions are also the symmetries of the given system of ODEs. If a given system of ODEs is not self-adjoint, then its integrating factors are necessarily solutions of the adjoint system of its linearized system. These solutions are known as adjoint-symmetries of the given system of ODEs. 

The adjoint ODE of linearized symmetry condition (\ref{met16}) can be written as (Bluman \& Anco, 2002)
\begin{equation}
L^*[x]\Lambda(t,x,\dot{x},\ddot{x})\equiv D^3[\Lambda]+D^2[\phi_{\ddot{x}}\Lambda]-D[\phi_{\dot{x}}\Lambda]+\phi_{x}\Lambda=0.\label{met23}
\end{equation}
where $D$ is the total derivative operator which is given by $D=\frac{\partial} {\partial t}+\dot{x}\frac{\partial} {\partial x}+\ddot{x}\frac{\partial} {\partial \dot{x}}+\phi \frac{\partial} {\partial \ddot{x}}$. Solutions of the above equation are known as adjoint-symmetries.
The adjoint-symmetry of the Eq.(\ref{main1}) becomes an integrating factor of (\ref{main1}) if and only if $\Lambda(t,x,\dot{x})$ satisfies the adjoint invariance condition
\begin{equation}
\Lambda_{t\ddot{x}}+\Lambda_{x\ddot{x}}\dot{x}+2\Lambda_{\dot{x}}+\ddot{x}\Lambda_{\dot{x}\ddot{x}}+(\phi \Lambda)_{\ddot{x}\ddot{x}} =0.\label{adjo1}
\end{equation}
Once we know the integrating factors, we can find the first integrals. Multiplying the given equation by these integrating factors, and rewriting the resultant equation as a perfect differentiable function, 
\begin{equation}
\Lambda(t,x,\dot{x},\ddot{x})(\dddot{x}-\phi(t,x,\dot{x},\ddot{x}))=\frac{d} {dt}\big(I\big),\label{adjint}
\end{equation}
we can identify the first integrals.

\subsection{$\lambda$-symmetries}
All the nonlinear ODEs do not necessarily admit Lie point symmetries.  Under such a circumstance one may look for  generalized symmetries (other than Lie point symmetries) associated with the given equation.  One such generalized symmetry is the $\lambda$-symmetry. Let $\tilde{V}=\xi(t,x)\frac{\partial}{\partial t}+\eta(t,x)\frac{\partial}{\partial x}$ be a $\lambda$-symmetry of the given ODE for some function $\lambda=\lambda(t,x,\dot{x},\ddot{x})$. The invariance of the given ODE (\ref{main1}) under $\lambda$-symmetry vector field is given by $\tilde{V}^{[\lambda,(3)]}(\dddot{x}-\phi(t,x,\dot{x},\ddot{x}))=0$, where $\tilde{V}^{[\lambda,(3)]}$ is given by $\xi \frac{\partial} {\partial t}+\eta \frac{\partial} {\partial x}+\eta^{[\lambda,(1)]}\frac{\partial} {\partial \dot{x}}+\eta^{[\lambda,(2)]}\frac{\partial} {\partial \ddot{x}}+\eta^{[\lambda,(3)]}\frac{\partial} {\partial \dddot{x}}$. Here $\eta^{[\lambda,(1)]}$, $\eta^{[\lambda,(2)]}$ and $\eta^{[\lambda,(3)]}$ are first, second and third $\lambda$- prolongations respectively whose explicit expressions are given by (Muriel \& Romero, 2001, 2008, 2009)
\begin{eqnarray}
\eta^{[\lambda,(1)]}&=&(D+\lambda)\eta(t,x)-(D+\lambda)(\xi(t,x))\dot{x},\label{etlam1} \nonumber\\
\eta^{[\lambda,(2)]}&=&(D+\lambda)\eta^{[\lambda,(1)]}(t,x,\dot{x})-(D+\lambda)(\xi(t,x))\ddot{x},\label{etlam2}\nonumber \\
\eta^{[\lambda,(3)]}&=&(D+\lambda)\eta^{[\lambda,(2)]}(t,x,\dot{x},\ddot{x})-(D+\lambda)(\xi(t,x))\dddot{x}.\label{etlam3}
\end{eqnarray}
Expanding the invariance condition, we find
\begin{equation}
\xi\phi_t+\eta\phi_x+\eta^{[\lambda,(1)]}\phi_{\dot{x}}+\eta^{[\lambda,(2)]}\phi_{\ddot{x}}-\eta^{[\lambda,(3)]}=0,
\label{beq1}
\end{equation}
where the infinitesimal prolongations are as given above.
If we put $\lambda=0$, we can get the Lie prolongation formula. Solving the invariance condition (\ref{beq1}) we can obtain the explicit forms of $\xi, \eta$ and $\lambda$.
If the given ODE admits Lie point symmetries, then the $\lambda$-symmetries can be derived without solving the $\lambda$-prolongation condition. In this case the $\lambda$-symmetries can be deduced from the relation
\begin{equation}
\lambda=\frac{D[Q]} {Q},\label{qlam}
\end{equation}
where $D$ is the total differential operator and $Q=\eta-\xi\dot{x}$. The associated vector field is given by $\tilde{V}=\frac{\partial} {\partial x}$. The method of finding the integrals from $\lambda$-symmetries can also be extended to the third-order ODEs as in the case of second-order ODEs. 

\subsection{Extended Prelle-Singer method}
In a series of papers Chandrasekar, Senthilvelan and Lakshmanan have developed a stand-alone method, namely extended Prelle-Singer procedure, to investigate the integrability of the given ODE which may be of any order, including coupled ones. In the following, we recall briefly the extended Prelle-Singer procedure which is applicable for third-order ODEs (Chandrasekar \textit{et al.} 2006). 

We assume that the ODE (\ref{main1}) admits a first integral $I(t,x,\dot{x},\ddot{x})=C,$ with $C$ constant on the solutions, so that the total differential gives 
\begin{equation}  
dI={I_t}{dt}+{I_{x}}{dx}+{I_{\dot{x}}{d\dot{x}}}+{I_{\ddot{x}}{d\ddot{x}}}=0, 
\label{met3}  
\end{equation}
where the subscript denotes partial differentiation with respect to that variable.  Rewriting equation~(\ref{main1}) in the form 
$\frac{P}{Q}dt-d\ddot{x}=0$ and adding the null terms $U(t,x,\dot{x},\ddot{x})\ddot{x}dt - U(t,x,\dot{x},\ddot{x})d\dot{x}$ and
$S(t,x,\dot{x},\ddot{x})\dot{x}dt - S(t,x,\dot{x},\ddot{x})dx$ to the latter, we obtain that on 
the solutions the 1-form
\begin{equation}
\bigg(\frac{P}{Q}+S\dot{x}+U\ddot{x}\bigg) dt-Sdx-Ud\dot{x}-d\dot{x} = 0. 
\label{met6} 
\end{equation}	
Hence, on the solutions, the 1-forms (\ref{met3}) and (\ref{met6}) must be proportional, provided (\ref{met6}) is a total differential. To ensure this we multiply (\ref{met6}) by the function $ R(t,x,\dot{x},\ddot{x})$ which acts as the integrating factor for (\ref{met6}) so that we have on the solutions that 
\begin{equation} 
dI=R(\phi+S\dot{x}+U\ddot{x})dt-RSdx-RUd\dot{x}-Rd\ddot{x}=0.
\label{met7}
\end{equation}
Comparing now equations (\ref{met3}) 
with (\ref{met7}) we end up with a set of four relations which relates the integral($I$), integrating factor($R$) and the null term($S$): 
\begin{equation} 
I_{t}=R(\phi+\dot{x}S+U\ddot{x}),~I_{x}=-RS,~I_{\dot{x}}=-RU,~I_{\ddot{x}}=-R.  
\label{met8}
\end{equation} 

In order to determine $S, U$ and $R$, we impose the compatibility conditions $I_{tx}=I_{xt}$, $I_{t\dot{x}}=I_{{\dot{x}}t}$, $I_{t\ddot{x}}=I_{{\ddot{x}}t}$,
$I_{x{\dot{x}}}=I_{{\dot{x}}x}$, $I_{x{\ddot{x}}}=I_{{\ddot{x}}x}$ and $I_{\dot{x}{\ddot{x}}}=I_{{\ddot{x}}\dot{x}}$. Then we obtain the following determining equations, 
\begin{eqnarray}  
D[S]=&-\phi_x+S\phi_{\ddot{x}}+US,\label{met9}\\
D[U]=&-\phi_{\dot{x}}+U\phi_{\ddot{x}}-S+U^2,\label{met91}\\
D[R]=&-R(U+\phi_{\ddot{x}}),\label{met10}\\
R_x=&R_{\ddot{x}}S+RS_{\ddot{x}},\label{met11}\\
R_{\dot{x}}S=&-S_{\dot{x}}R+R_x U+RU_x,\label{met111}\\
R_{\dot{x}}=&R_{\ddot{x}}U+RU_{\ddot{x}},\label{met112}
\end{eqnarray}
where $D=\frac{\partial}{\partial{t}}+
\dot{x}\frac{\partial}{\partial{x}}+
\ddot{x}\frac{\partial}{\partial{\dot{x}}}+\phi\frac{\partial}
{\partial{\ddot{x}}}$.
Solving the above system of over determined equations we can obtain the unknown functions $S,~U$ and $R$. From the known expressions, $S,~ U$ and $R$, we can determine the integrals which appear on the left hand side of Eq.(\ref{met8}) by straightforward integration. 

Integrating Eq.(\ref{met8}) we find
\begin{eqnarray}
I(t,x,\dot{x},\ddot{x})&=&r_1-r_2-\int\left[U+\frac{d}{d \dot{x}}[r_1-r_2]\right]d\dot{x}\nonumber\\&&-\int \left[R+\frac{d}{d\ddot{x}}[r_1-r_2-\int [RU+\frac{d}{d\dot{x}}[r_1-r_2]]d\dot{x}]\right]d\ddot{x}, \label{met13}
\end{eqnarray}
where
\begin{equation}
 r_1 =\int R(\phi+S\dot{x}+U\ddot{x})dt,~~
 r_2 = \int \left(RS+\frac{d}{dx}\int r_1\right)dx.\nonumber
\end{equation}

We may note that every independent set $(S, U, R)$ in (\ref{met13}) defines a first integral.

\section{Interconnections}

In the previous section we have discussed six specific analytical methods which are used to derive integrating factors, symmetries of various kinds, null forms and integrals associated with the third-order ODEs. A question which we raise now is what is the interconnection between these various methods, that is given any one of the quantities in the family, say multipliers, Darboux polynomials, Lie point symmetries, adjoint-symmetries, $\lambda$-symmetries or integrating factors and null forms, can one obtain the rest of the quantities in this family. To answer this question we can explore the hidden interconnections that exist between these functions and interlink all these methods. To achieve this task we introduce certain transformations in the extended Prelle-Singer procedure which in turn connect globally all the above mentioned quantities. The details are given below (It may be noted that one can establish the same interconnection by taking any one of the other methods as the starting point).

With the aid of the following transformations on the null forms of the PS method, see Eq.(\ref{met9}),
\begin{equation}
U=-\frac{D[V]}{V}~~\mathrm{and}~~S=\frac{X}{V},\label{ueq}
\end{equation}
where $V(t,x,\dot{x},\ddot{x})$ and $X(t,x,\dot{x},\ddot{x})$ are two unknown functions and $D$ is the total differential operator, Eqs.(\ref{met91}) and (\ref{met9}) respectively can be rewritten in the form
\begin{equation}
D^2[V]=D[V]\phi_{\ddot{x}}+V\phi_{\dot{x}}+X,\label{veq}
\end{equation}
and
\begin{equation}
D[X]=\phi_{\ddot{x}} X-\phi_x V.\label{met14}
\end{equation}
We introduce another transformation on the integrating factor as
\begin{equation}
R=V/F,\label{rxf}
\end{equation}
where $F(t,x,\dot{x},\ddot{x})$ is a function to be determined, in (\ref{met10}) so that the latter can be rewritten in a compact form in the new variable $F$ as
\begin{equation}  
D[F] = \phi_{\ddot{x}}F.
\label{met15}
\end{equation}
One may note that Eq.(\ref{met15}) is nothing but the Darboux polynomial determining equation (\ref{Darb}) with the cofactor $g=\phi_{\ddot{x}}$. 

We mention here that once we know the functions $U$ and $S$ the integrating factor $R$ can be determined within the Prelle-Singer procedure itself. But to connect the integrating factors to Darboux polynomials these transformations are essential. More importantly the connection between the null forms and $\lambda$-symmetries can be unearthed through the function $V$ which appears in the transformations (\ref{ueq}) and (\ref{rxf}), as we see below.

\subsection{Connection between $\lambda$-symmetries and null forms}
Now we investigate how these expressions, namely $U$ and $S$ are interconnected with $\lambda$-symmetries. In the case of second-order ODEs the $\lambda$-symmetry is nothing but the null form with a negative sign (Muriel \& Romero, 2009). This one-to-one correspondence came from the result that the $S$-determining equation in the PS procedure differs only by a negative sign from that of the $\lambda$-symmetry determining equation (Mohanasubha \textit{et al.} 2014). However, in the case of third-order ODEs we have two null forms $S$ and $U$ which have to be connected to a single function $\lambda$ as demonstrated below. 

Let $I(t,x,\dot{x},\ddot{x})$ be a first integral of (\ref{main1}) then $R=-I_{\ddot{x}}$ is an integrating factor, see Eq.(\ref{met8}). The total derivative $\frac{dI} {dt}=0$ gives
\begin{equation}
 R \phi=I_t+\dot{x} I_x+\ddot{x}I_{\dot{x}}.
\end{equation}
Let $I(t,x,\dot{x},\ddot{x})$ also be a first integral of $\tilde{V}^{[\lambda,(2)]}$ for some function $\lambda(t,x,\dot{x},\ddot{x})$, then 
\begin{equation}
\tilde{V}^{[\lambda,(2)]}I=0,\label{exlam}
\end{equation}
where $\tilde{V}^{[\lambda,(2)]}$ is given by $\xi \frac{\partial} {\partial t}+\eta \frac{\partial} {\partial x}+\eta^{[\lambda,(1)]}\frac{\partial} {\partial \dot{x}}+\eta^{[\lambda,(2)]}\frac{\partial} {\partial \ddot{x}}$. Here $\eta^{[\lambda,(1)]}$ and $\eta^{[\lambda,(2)]}$ are the first and second $\lambda$- prolongations, respectively, whose explicit expressions are given in the first two expressions of (\ref{etlam3}). Expanding Eq. (\ref{exlam}) we get the following expression for $\lambda$ corresponding to $\tilde{V}=\xi\frac{\partial}{\partial t}+\eta\frac{\partial}{\partial x}$,
\begin{eqnarray}
\xi \frac{\partial I} {\partial t}+\eta \frac{\partial I} {\partial x}+( \eta^{(1)}+\lambda(\eta-\xi \dot{x}))\frac{\partial I} {\partial \dot{x}}+(\eta^{(2)}+D[\lambda](\eta-\xi \dot{x})\nonumber \\+2\lambda(\eta^{(1)}-\xi\ddot{x})+\lambda^2(\eta-\xi \dot{x}))\frac{\partial I} {\partial \ddot{x}}=0,\label{expan}
\end{eqnarray}
where $\eta^{(1)}$ and $\eta^{(2)}$ are the first and second Lie point prolongations which are given in Eq.(\ref{prol}). For the vector field $\tilde{V}=\frac{\partial}{\partial x}$ the above expression (\ref{expan}) reduces to
\begin{eqnarray}
-(D[\lambda]+\lambda^2)I_{\ddot{x}}=I_x+\lambda I_{\dot{x}}.\label{ejg}
\end{eqnarray}

Now we connect this expression which comes out from the $\lambda$-symmetry analysis with the null forms in the extended PS procedure. In this regard we can deduce the following expressions from the last three equations of (\ref{met8}),
\begin{equation}
S=\frac{I_x} {I_{\ddot{x}}},~~U=\frac{I_{\dot{x}}}{I_{\ddot{x}}},~~I_{\ddot{x}} \neq 0.\label{ejg1}
\end{equation}
Using the above in Eq. (\ref{ejg}) we can obtain an equation which connects the null forms $U$ and $S$ with $\lambda$ in the form
\begin{equation}
U=-\frac{(D[\lambda]+\lambda^2+S)} {\lambda}.\label{lamus}
\end{equation}
Eq. (\ref{lamus}) is essentially a redefinition of the $\lambda$-function determining equation (\ref{ejg}). Thus in the case of third-order ODEs the null forms, $U$ and $S$, are connected with the  $\lambda$-symmetries through a differential relation. This interconnection is brought out for the first time in the literature. We mention here that while deriving the relation (\ref{lamus}) we assumed that $\lambda \neq 0$. When $\lambda =0$ we have $I_x=0$ (vide Eq.(\ref{ejg})) and in this case one of the null forms $(S)$ vanishes. This result is also consistent with the extended PS procedure. 

For practical purpose we can rewrite the relation (\ref{lamus}) in terms of a single variable, say in $V$, as follows. Using (\ref{veq}) we can express $X$ in terms of $V$ and substituting the latter in the second expression in (\ref{ueq}) we find $S$ in terms of $V$. The resultant expression reads
\begin{eqnarray}
S=\frac{1}{V}(D^2[V]-\phi_{\ddot{x}}D[V]-\phi_{\dot{x}}V).\label{sfg}
\end{eqnarray}
Now replacing the null functions, $S$ and $U$, which appear in (\ref{lamus}) by $V$, we find 
\begin{eqnarray}
D^2[V]-(\phi_{\ddot{x}}+\lambda)D[V]+(D[\lambda]+\lambda^2-\phi_{\dot{x}})V=0.\label{veqn1}
\end{eqnarray}
Substituting the known expressions $\phi_{\dot{x}}$ and $\phi_{\ddot{x}}$ and $\lambda$ in (\ref{veqn1}) and solving the resultant equation we can obtain $V$ which in turn unambiguously fixes the null forms $U$ and $S$ through the relations given in Eq.(\ref{ueq}). In other words one can get the null forms $U$ and $S$ from $\lambda$ by finding the function $V$ also. In this sense Eq.(\ref{veqn1}) may also treated as a ``second bridge" which connects $\lambda$-symmetries with null forms. If we already know the null forms $U$ and $S$, Eq.(\ref{lamus}) yields the $\lambda$-symmetries in a straightforward manner. In this sense one can determine (i) $\lambda$ from $U$ and $S$ and (ii) $U$ and $S$ from $\lambda$. The expressions (\ref{lamus}) and (\ref{veqn1}) interconnect the PS procedure and the $\lambda$ symmetry analysis.

\subsection{Connection between Lie point symmetries and null forms}
The relation between Lie point symmetries and $\lambda$-symmetries has already been established earlier (vide Eq.(\ref{qlam})). Substituting this in (\ref{lamus}) we can obtain an expression that relates Lie point symmetries with the null forms in the following manner:
\begin{eqnarray}
D^2[Q]+U D[Q]+S Q=0,
\end{eqnarray}
where $Q=\eta-\dot{x}\xi$ is the characteristics.

For the sake of completeness, in the following, we recall the following interconnection that is already known in the literature.
\subsection{Connection between Jacobi last multiplier and Darboux polynomials/integrating factors}
Using the expression (\ref{rxf}) we can relate the Darboux polynomials with the integrating factor and in fact  the denominator of (\ref{rxf}) is nothing but the Darboux polynomials. 

By comparing Eqs.(\ref{Darb}) and (\ref{met20}) we can relate the Darboux polynomials with Jacobi last multiplier as 
\begin{equation}                  
M=F^{-1}.
\end{equation}
Using this relation, we can find the Jacobi last multiplier from the Darboux polynomials. So the integrating factor $R$ in the PS method is the product of the function $V$ and the Jacobi last multiplier $M$, that is $R=VM$.

\subsection{Connection between adjoint-symmetries and integrating factors}
We rewrite Eqs.(\ref{met9}),~~(\ref{met91}) and (\ref{met10}) as a single equation in one variable, for example in $R$. Then the resultant equation reads
\begin{equation}  
D^3[R]+D^2[\phi_{\ddot{x}}R]-D[\phi_{\dot{x}}R]+\phi_{x}R=0.\label{met24}
\end{equation}
Comparing the adjoint of the linearized symmetry condition equation (\ref{met23}) and (\ref{met24}) one can conclude that the integrating factor $R$ is nothing but the adjoint-symmetry $\Lambda$, that is 
\begin{equation}
R=\Lambda. \label{met25}
\end{equation}
Thus the integrating factor turns out to be the adjoint-symmetry of the given third-order nonlinear ODE.

\subsection{Connection between Lie point symmetries and Jacobi last multipliers}
The connection between Lie point symmetries and JLM is known for a long time in the form 
\begin{equation}
 M=\frac{1} {\Delta},
\end{equation}
where
\begin{equation}
\Delta = \begin{vmatrix}
1 & \dot{x} & \ddot{x} & \dddot{x} \\
\xi_1 & \eta_1 & \eta_{1}^{(1)} & \eta_{1}^{(2)} \\
\xi_2 & \eta_2 & \eta_{2}^{(1)} & \eta_{2}^{(2)} \\
\xi_3 & \eta_3 & \eta_{3}^{(1)} & \eta_{3}^{(2)}\end{vmatrix},\label{delta}
\end{equation}
where $(\xi_1,\eta_1)$, $(\xi_2,\eta_2)$ and $(\xi_3,\eta_3)$  are three sets of Lie point symmetries (see below) of the third-order ODE, $\eta_{1}^{(i)}$, $\eta_{2}^{(i)}$ and $\eta_{3}^{(i)}$, $i=1,2$, are their corresponding first and second prolongations, respectively, and the inverse of $\Delta$ becomes the multiplier of the given equation (Nucci, 2005).

\subsection{Comparison between interconnections for second- and third-order ODEs}
In the above said interconnections, some of the interconnections are common to both second- and third-order nonlinear ODEs except for their orders, while the others are different. Such common and differing connections are listed below.\vspace{0.15cm}\\
1) {\it Common features:}\vspace{0.15cm}\\
The common connections are the ones between (i) Lie point symmetries and $\lambda$-symmetries, (ii) Lie point symmetries and Jacobi last multiplier, (iii) adjoint-symmetries and integrating factors, and (iv) Darboux polynomials and Jacobi last multiplier.\vspace{0.15cm}\\ 
2) {\it Differing features:}\vspace{0.15cm}\\
(i) The uncommon relation is the connection between $\lambda$-symmetries and null forms. In the cases of second- and third-order ODEs the connection between $\lambda$-symmetries and null forms are entirely different. In the case of second-order ODEs the $\lambda$-symmetry is nothing but the null form with a negative sign. This one-to-one correspondence came from the result that the $S$-determining equation in the PS procedure differs only by a negative sign from that of the $\lambda$-symmetry determining equation (Muriel \& Romero, 2009). In other words there is an one-to-one correspondence between the $\lambda$-symmetries and null form $S$. However, in the case of third-order ODEs, we have two null forms ($S$ and $U$) which have to be connected to the single function $\lambda$. Our analysis shows that these two null forms are connected with the $\lambda$-symmetry through a single expression.\vspace{0.2cm} \\
(ii) The connection between characteristics and null forms is also different in the case of second- and third-order ODEs. In the case of second-order ODEs, there exists only one null form which is directly connected with the characteristics, while in the case of third-order ODEs, the equation which connects the characteristics and null forms contains both the null forms. 

\subsection{Flow chart of the interconnections}
The above interconnections are clearly depicted in Fig.1. This may be compared with that of the second-order ODEs given as Fig.1 in Mohanasubha \textit{et al.} (2014).
\begin{figure}[h!]
  \centering
   \includegraphics[width=0.70\textwidth]{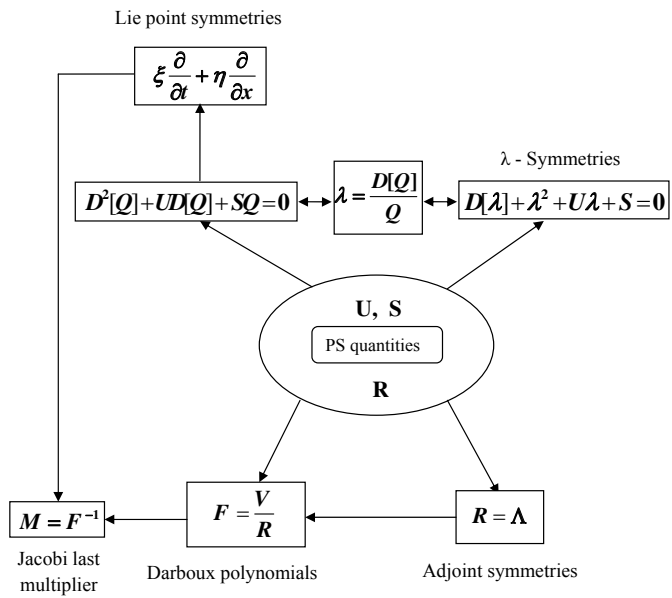}
    \caption{Flow chart connecting Prelle-Singer procedure with other methods for third-order ODEs}
\end{figure}

\section{Illustration}
\subsection{Example 1:}
We consider the following third-order nonlinear ODE, namely
\begin{equation}
\dddot{x}-\frac{\ddot{x}^2}{\dot{x}}-\frac{\dot{x}\ddot{x}}{x}=0.\label{exam}
\end{equation}
Equation (\ref{exam}) is a sub-case of the general form of a scalar third-order ODE which is invariant under the generators of time translation and rescaling (Feix \textit{et al.} 1997; Polyanin \& Zaitsev 2003). A sub-case of equation (\ref{exam}) has been considered by both Bocharov \textit{et al.} (1993) and Ibragimov \& Meleshko (2005) in the form $\dddot{x}-c\frac{\ddot{x}^2}{\dot{x}}=0$. They showed that it can be linearized to a linear third-order ODE through a contact transformation. The equation $\dddot{x}-\frac{\dot{x}\ddot{x}}{x}=0$ has been considered by Euler \& Euler (2004), who showed that it can be linearized to a free particle equation through a Sundman transformation. Equation (\ref{exam}) was studied by Chandrasekar \textit{et al.} (2005) from the integrability point of view using Prelle-Singer procedure. Here we consider Eq.(\ref{exam}) from the perspective of deriving multipliers, Darboux polynomials, Lie point symmetries, adjoint-symmetries, $\lambda$-symmetries, null forms, integrating factors, sequentially and demonstrate the effectiveness of exploring the interconnections. From these quantities we also derive integrals and the general solution of this equation for the sake of completeness. 

We begin our analysis with Lie point symmetries. Equation (\ref{exam}) admits a set of three parameter Lie point symmetries of the form
\begin{eqnarray}
\Omega_1=\frac{\partial}{\partial t}, \quad \Omega_2=-x\frac{\partial}{\partial x},\quad \Omega_3=t\frac{\partial}{\partial t}.
\end{eqnarray}
The characteristics associated with the above vector fields is given by 
\begin{eqnarray}
Q_1=-\dot{x},\qquad Q_2=-x,\qquad Q_3=-t\dot{x}.
\end{eqnarray}
Using the relation $\lambda=\frac{D[Q]}{Q}$ we can derive the $\lambda$-symmetries associated with the above vector fields which in turn read
\begin{equation}
\lambda_1=\frac{\ddot{x}}{\dot{x}}, ~~\lambda_2=\frac{\dot{x}}{x},~~\lambda_3=\frac{\dot{x}+t\ddot{x}}{t\dot{x}}.\label{lamex3}
\end{equation}
Once we know the $\lambda$-symmetry we can find the null forms $U$ and $S$. To obtain them we determine the function $V$ using the relation (\ref{veqn1}). We obtain
\begin{equation}
 V_1=\dot{x},~~~V_2=\dot{x}^2-x\ddot{x},~~~V_3=t\dot{x}^2-x(\dot{x}+t\ddot{x}).\label{vexa}
\end{equation}
We can fix the allowed forms of the function $X$ from the known expression of $V$ through the relation (\ref{veq}). The resultant expressions read
\begin{equation}
X_1= \frac{\dot{x}\ddot{x}}{x},~~~X_2=-\frac{\dot{x}^2\ddot{x}}{x},~~~X_3=-\frac{\dot{x}\ddot{x}(x+t\dot{x})}{x}.\label{xexa}
\end{equation}
Using the functions $X$ and $V$ we can obtain the null forms $U$ and $S$ respectively, vide Eq.(\ref{ueq}), as
\begin{equation}
S_1=-\frac{\ddot{x}} {x},~~S_2=-\frac{\dot{x}^2\ddot{x}} {x(\dot{x}^2-x\ddot{x})},~~S_3=\frac{\dot{x}\ddot{x}(x+t\dot{x})} {x(-t\dot{x}^2+x(\dot{x}+t\ddot{x}))},\label{ssexa}
\end{equation}
and 
\begin{equation}
U_1=-\frac{\ddot{x}} {\dot{x}},~~U_2=\frac{x\ddot{x}^2} {\dot{x}(\dot{x}^2-x\ddot{x})},~~U_3=\frac{\dot{x}\ddot{x}(2\dot{x}+t\ddot{x})} {\dot{x}(t\dot{x}^2-x(\dot{x}+t\ddot{x}))}.\label{uuexa}
\end{equation}
Substituting the null forms $U$ and $S$ in Eq.(\ref{met10}) and solving the resultant equation we find the integrating factors for (\ref{exam}) in the form
\begin{equation}
R_1=\frac{-1} {\dot{x}x},~~
R_2=\frac{2\ddot{x}} {\dot{x}^2}-\frac{2}{x},~~
R_3=\frac{t\dot{x}^2-x(\dot{x}+t\ddot{x})}{2 x \dot{x}\sqrt{\frac{\ddot{x}(2\dot{x}^2-x\ddot{x})}{x}}}.\label{req3}
\end{equation}
The functions $R_1,R_2$ and $R_3$ also become adjoint-symmetries of (\ref{exam}) as well which can be directly verified by substituting them in (\ref{met23}).
Using the expressions for $R$ and $V$ we can fix the Darboux polynomials for the given equation by recalling the relation $F=\frac{V}{R}$. The Darboux polynomials turn out to be
\begin{equation}
F_{11}=-x\dot{x}^2,~~
F_{12}=-\frac{x\dot{x}^2}{2},~
F_{13}=2x\dot{x}^2.\label{dp3v}
\end{equation}
The polynomials $F_{11},F_{12}$ and $F_{13}$ share the common cofactor $\frac{\dot{x}}{x}+\frac{2\ddot{x}}{\dot{x}}$. From the property that the ratio of Darboux polynomials which share the same cofactor is the first integral, we can find another Darboux polynomial which shares the cofactor as the above. Doing so, we have found two more Darboux polynomials satisfying the equation $D[F]=\big(\frac{\dot{x}}{x}+\frac{2\ddot{x}}{\dot{x}}\big)F$ as
\begin{equation}
F_2=\ddot{x}(2\dot{x}^2-\ddot{x}x),~~F_3=\dot{x}\ddot{x}.\label{da4}
\end{equation}
We can find more number of Darboux polynomials using the property of Darboux polynomial that the combination of Darboux polynomials is also a Darboux polynomial. By multiplying an integral with the Darboux polynomial we can obtain more Darboux polynomials. The ratios of the Darboux polynomials $\frac{F_3} {F_1}$ and $\frac{F_2} {F_1}$ lead to the first integrals $I_1$ and $I_2$ (vide Eq.(\ref{int1})), respectively.

The JLM associated with the given equation can be obtained from the Darboux polynomials (\ref{dp3v}) by recalling the relation $F=M^{-1}$.

Once we know the null forms and integrating factors we can construct the associated integrals of motion by evaluating the integrals given in (\ref{met13}). Our analysis shows that
\begin{equation}
I_1=\frac{\ddot{x}} {\dot{x}x},~ 
I_2=\ddot{x}\bigg(\frac{2} {x}-\frac{\ddot{x}} {\dot{x}^2}\bigg),~I_3=-\frac{t} {2\dot{x}}\sqrt{\frac{\ddot{x}} {x}(2\dot{x}^2-x\ddot{x})}+\tan^{-1}\sqrt{\frac{\ddot{x}} {x(2\dot{x}^2-x\ddot{x})}}x.\label{int1} 
\end{equation}
It is a straightforward matter to verify that $I_1$, $I_2$ and $I_3$ are the three independent integrals for the given Eq.(\ref{exam}). From the knowledge of $I_1$, $I_2$ and $I_3$ we get the general solution in the form
\begin{equation}
x(t)=\frac{\sqrt{I_2}} {I_1}\tan\bigg[\frac{\sqrt{I_2}} {2}t+I_3\bigg].\label{soln}
\end{equation}
The above said results are tabulated in Table 1.

\subsection{Example 2:}
We consider another interesting example (Bluman \& Anco, 2002)
\begin{equation}
\dddot{x}=\frac{6t\ddot{x}^3} {\dot{x}^2}+\frac{6\ddot{x}^2} {\dot{x}}.\label{exam1}
\end{equation}
The adjont symmetries of this equation were first worked out by Bluman and Anco (2002). Subsequently the extended Prelle-Singer method was applied to this equation by Chandrasekar \textit{et al.} (2005) and they have derived the null forms, integrating factors and integrals through this method. The exact expressions are given in Table 2. The procedure given in the previous example may be followed for this example as well to derive the quantities displayed in Table 2. Here also because of nonuniqueness of $S,U$ with $\lambda$, one may find two independent integrals $I_2$ and $I_3$ from the same $\lambda$-function. The vector field $\Omega_1$ provides one independent integral $I_1$ and the vector field $\Omega_2$ gives the two independent integrals $I_2$ and $I_3$. The vector field $\Omega_3$ provides only a dependent integral.

\begin{sidewaystable}[ht]
\small\addtolength{\tabcolsep}{-3pt}
{
\newcommand{\mc}[3]{\multicolumn{#1}{#2}{#3}}
\begin{center}
\begin{tabular}{|l|c|c|c|c|c|c|c|c|} \hline
\hspace{0.3cm}$\Omega$ & $\lambda$ & $V$ & $U$ & $X$ & $S$ & $F$ & $R$ & $I$\\\hline
\mc{1}{|c|}{$\frac{\partial}{\partial t}$} & $\frac{\ddot{x}}{\dot{x}}$ & $\dot{x}$ & $-\frac{\ddot{x}} {\dot{x}}$   & $-\frac{\dot{x}\ddot{x}}{x}$ & $ -\frac{\ddot{x}} {x}$ & $-x\dot{x}^2 $ & $\frac{-1} {x\dot{x}}$  & $ I_1=\frac{\ddot{x}} {x\dot{x}}$ \\\cline{3-9}
\mc{1}{|c|}{ } & $$ & $x^2\dot{x}$ & $-\bigg(\frac{2\dot{x}} {x}+\frac{\ddot{x}} {\dot{x}}\bigg)$ & $x\dot{x}\ddot{x}$ & $\frac{\ddot{x}}{x}$ & $ x\dot{x}^2$ & $\frac{x} {\dot{x}}$ & $I_4=\frac{I_2}{I_1}$  \\\hline

\mc{1}{|c|}{$-x\frac{\partial}{\partial x}$} & $\frac{\dot{x}}{x}$ & $\dot{x}^2-x\ddot{x}$ & $\frac{x\ddot{x}^2} {\dot{x}(\dot{x}^2-x\ddot{x})}$ & $-\frac{\dot{x}^2\ddot{x}}{x}$ & $-\frac{\dot{x}^2\ddot{x}} {x(\dot{x}^2-x\ddot{x})} $ & $ \frac{-x\dot{x}^2}{2}$ & $ \frac{2\ddot{x}} {\dot{x}^2}-\frac{2}{x}$ & $I_2=\ddot{x}\bigg(\frac{2} {x}-\frac{\ddot{x}} {\dot{x}^2}\bigg)$\\\hline

\mc{1}{|c|}{$t\frac{\partial}{\partial t}$} & $\frac{\dot{x}+t\ddot{x}}{t\dot{x}}$ & $t\dot{x}^2-x(\dot{x}+t\ddot{x})$ & $\frac{\dot{x}\ddot{x}(2\dot{x}+t\ddot{x})} {\dot{x}(t\dot{x}^2-x(\dot{x}+t\ddot{x}))}$  & $-\frac{\dot{x}\ddot{x}(x+t\dot{x})}{x}$& $\frac{\dot{x}\ddot{x}(x+t\dot{x})} {x(-t\dot{x}^2+x(\dot{x}+t\ddot{x}))}$ & $2 x \dot{x}\sqrt{\frac{\ddot{x}(2\dot{x}^2-x\ddot{x})}{x}} $ & $ \frac{t\dot{x}^2-x(\dot{x}+t\ddot{x})}{2 x \dot{x}\sqrt{\frac{\ddot{x}(2\dot{x}^2-x\ddot{x})}{x}}}$ & $I_3=-\frac{t} {2\dot{x}}\sqrt{\frac{\ddot{x}} {x}(2\dot{x}^2-x\ddot{x})}$ \\
& & & & & & & & $+\tan^{-1}\sqrt{\frac{\ddot{x}} {x(2\dot{x}^2-x\ddot{x})}}x$ \\
\hline

\end{tabular}
\caption{Vector fields $(\Omega)$, $\lambda$-symmetries, null forms $(U,S)$, Darboux polynomials $(F)$, integrating factors $(R)$ and integrals $(I)$ admitted by Eq. (\ref{exam}), along with the quantities $V$ and $X$ defined by Eqs.(\ref{veq}) and (\ref{met14}), respectively.}
\end{center}
}%
\smallskip
\small\addtolength{\tabcolsep}{-3pt}
{
\newcommand{\mc}[3]{\multicolumn{#1}{#2}{#3}}
\begin{center}
\begin{tabular}{|l|c|c|c|c|c|c|c|c|} \hline
\hspace{0.3cm}$\Omega$ & $\lambda$ & $V$ & $U$ & $X$ & $S$ & $F$ & $R$ & $I$\\\hline

\mc{1}{|c|}{$t\frac{\partial}{\partial t}$} & $\frac{\dot{x}+t\ddot{x}}{t\dot{x}}$ & $\frac{\ddot{x}}{\dot{x}^4}$ & $-\frac{2\dot{x}\ddot{x}+6t\ddot{x}^2} {\dot{x}^2}$   & $\frac{2\ddot{x}^3}{\dot{x}^6}$ & $ \frac{2\ddot{x}^2} {\dot{x}^2}$ & $\bigg(\frac{\ddot{x}}{\dot{x}^2}\bigg)^3$ & $\frac{\dot{x}^2} {\ddot{x}^2} $ & $I_1=6t\dot{x}-2x+\frac{\dot{x}^2} {\ddot{x}}$\\\hline

\mc{1}{|c|}{$\frac{\partial}{\partial x}$} & $0$ & $\frac{\ddot{x}}{\dot{x}^3}$ & $-\frac{3\dot{x}\ddot{x}+6t\ddot{x}^2} {\dot{x}^2}$ & $0$ & $0 $ & $ \bigg(\frac{\ddot{x}}{\dot{x}^2}\bigg)^3 $ & $ \frac{\dot{x}^3} {\ddot{x}^2}$ & $I_2=3t\dot{x}^2+\frac{\dot{x}^3} {\ddot{x}}$ \\\cline{3-9}
\mc{1}{|c|}{ } & $$ & $\frac{\ddot{x}}{\dot{x}^2}$ & $-\frac{4\dot{x}\ddot{x}+6t\ddot{x}^2} {\dot{x}^2}$ & $0$ & $0$ & $ \bigg(\frac{\ddot{x}}{\dot{x}^2}\bigg)^3$ & $ \frac{\dot{x}^4} {\ddot{x}^2}$ & $I_3=2t\dot{x}^3+\frac{\dot{x}^4} {\ddot{x}}$  \\\hline

\mc{1}{|c|}{$x\frac{\partial}{\partial x}$} & $\frac{\dot{x}}{x}$ & $\dot{x}$ & $-\frac{\ddot{x}}{\dot{x}}$  & $0$& $0$ & $\frac{3\ddot{x}^3} {\dot{x}}\bigg(\frac{\dot{x}+2t\ddot{x}}{\ddot{x}}\bigg)^{\frac{5} {3}}$ & $ \frac{\dot{x}^2} {3\ddot{x}^3\bigg(2t+\frac{\dot{x}} {\ddot{x}}\bigg)^{\frac{5} {3}}}$ & $\frac{I_2}{I_3{^{\frac{2}{3}}}}$ \\\hline

\end{tabular}
\caption{Vector fields $(\Omega)$, $\lambda$-symmetries, null forms $(U,S)$, Darboux polynomials $(F)$, integrating factors $(R)$ and integrals $(I)$ admitted by Eq. (\ref{exam1}), along with the quantities $V$ and $X$ defined by Eqs.(\ref{veq}) and (\ref{met14}), respectively}
\end{center}
}%
\end{sidewaystable}
Using the integrals $I_1, I_2$ and $I_3$ the solution of Eq.(\ref{exam1}) can be written down implicitly as
\begin{eqnarray}
&&3t(I_2(I_1-2x)+9tI_3)^2+I_2((I_1+2x)^2-12tI_2)^2\nonumber\\
&&-(I_1+2x)(I_2(I_1+2x)+9tI_3)((I_1+2x)^2-12tI_2)=0.
\end{eqnarray}

\subsection{Example 3:}
Finally, we consider the following example,
\begin{equation}
\dddot{x}=\frac{\dot{x} (\ddot{x}-1)}{x+1}.\label{exam3}
\end{equation}
Eq.(\ref{exam3}) admits two Lie point symmetries which are given by
\begin{equation}
 \Omega_1=\frac{\partial}{\partial t},~~\Omega_2=t\frac{\partial}{\partial t}+2(x+1)\frac{\partial}{\partial x}.\label{one_point}
\end{equation}
Since Eq.(\ref{exam3}) admits only two point symmetries, we start our analysis by exploring the null forms and the integrating factors in the PS method with the help of Eqs.(\ref{met9}), (\ref{met91}) and (\ref{met10}). Integrating these three equations, we obtain the following particular solutions for the null forms $S_i$ and $U_i,~i=1,2,3$, namely
\begin{eqnarray}
&&\hspace{-3.4cm}S_1=\frac{1-\ddot{x}}{x+1},~~~~S_2=\frac{x^2 (\ddot{x}+1)+2 x (\ddot{x}+1)+2 \ddot{x}}{x (x+1) (x+2)},\nonumber \\
&&\hspace{-3.4cm}S_3=\frac{(\ddot{x}-1) \left(t \left(\dot{x}^2 (\ddot{x}-1)-(x+1) \ddot{x}^2\right)+(x+1) \dot{x} \ddot{x}\right)}{(x+1) \left(t (x+1) \ddot{x}^2+\dot{x}^2 (t-t \ddot{x})+(x+1) \dot{x} (\ddot{x}-2)\right)},\nonumber 
\end{eqnarray}
\begin{equation}
U_1=0,~U_2=\frac{-2 (x+1) \dot{x}}{x (x+2)},~
U_3=\frac{2 (x+1) (1-\ddot{x}) \ddot{x}}{t (x+1) \ddot{x}^2+\dot{x}^2 (t-t \ddot{x})+(x+1) \dot{x} (\ddot{x}-2)}. 
\end{equation}
The associated integrating factors are found to be
\begin{equation}
R_1=\frac{-1}{x+1},~R_2=\frac{x (x+2)}{2 (x+1)},~R_3=\frac{-t (x+1) \ddot{x}^2+t \dot{x}^2 (\ddot{x}-1)-(x+1) \dot{x} (\ddot{x}-2)}{2 (x+1) \sqrt{\frac{1-\ddot{x}}{x+1}} \left((x+1) \ddot{x}^2-\dot{x}^2 (\ddot{x}-1)\right)}.
\end{equation}
Once we know the null forms $S$ and $U$, we can deduce the $\lambda$-symmetries by solving Eq.(\ref{lamus}). Doing so, we find
\begin{equation}
\lambda_1=\lambda_2=\frac{\ddot{x}}{\dot{x}},~~\lambda_3=\frac{\dot{x} (\ddot{x}-1)}{(x+1) \ddot{x}}.
\end{equation}
The null form pair ($S_3,U_3$) provides another $\lambda$-symmetry which is given by
\begin{equation}
 \tilde{\lambda}_2=\frac{\dot{x}-t\ddot{x}}{2+2x-t\dot{x}}.
\end{equation}
The $\lambda$-functions $\lambda_1(=\lambda_2)$ and $\tilde{\lambda}_2$ can also be obtained directly from the Lie point symmetries (\ref{one_point}).
In this example also, we observe that for a single $\lambda$ one can have multiple null forms. From the null forms, the functions $V$ and $X$ can be determined with the help of Eq. (\ref{ueq}) as
\begin{eqnarray}
V_1&=&1,~V_2=x (x+2),~V_3=\frac{t (x+1) \ddot{x}^2+\dot{x}^2 (t-t\ddot{x})+(x+1) \dot{x} (\ddot{x}-2)}{x+1},\nonumber \\
X_1&=&\frac{1-\ddot{x}}{x+1},~ X_2= \frac{x^2 (\ddot{x}+1)+2 x (\ddot{x}+1)+2 \ddot{x}}{x+1},\nonumber \\
X_3&=&\frac{(\ddot{x}-1) \left(t \left(\dot{x}^2 (\ddot{x}-1)-(x+1) \ddot{x}^2\right)+(x+1) \dot{x}\ddot{x}\right)}{(x+1)^2} .
\end{eqnarray}
Since the functions $V$ and $R$ are known now, the Darboux polynomials admitted by Eq.(\ref{exam3}) are found by exploiting the relation (\ref{rxf}). We observe
\begin{equation}
F_1=-(x+1),~~F_2=2 (x+1),~~F_3=-2 \sqrt{\frac{1-\ddot{x}}{x+1}} \left((x+1) \ddot{x}^2-\dot{x}^2 (\ddot{x}-1)\right).\label{3rd_dar}
\end{equation}
The JLM associated with the given equation can be obtained from the Darboux polynomials (\ref{3rd_dar}) by recalling the relation $F=M^{-1}$. Jaboci last multipliers are given by
\begin{equation}
M_1=-\frac{1} {1+x},~M_2=\frac{1}{2 x+2},~M_3=-\frac{1}{2 \sqrt{\frac{1-\ddot{x}}{x+1}} \left((x+1) \ddot{x}^2-\dot{x}^2 (\ddot{x}-1)\right)}.
\end{equation}
Once we know the null forms and integrating factors we can construct the associated integrals of motion by evaluating the integrals given in (\ref{met13}). Our analysis shows that
\begin{eqnarray}
I_1&=&\frac{x+\ddot{x}}{x+1},~~I_2=\frac{\dot{x}^2-x^2 (\ddot{x}+1)+x \left(\dot{x}^2-2 \ddot{x}\right)}{2 (x+1)},\nonumber \\
I_3&=&\tan ^{-1}\left(\frac{(x+1) \ddot{x} \sqrt{\frac{1-\ddot{x}}{x+1}}}{\dot{x} (\ddot{x}-1)}\right)-t \sqrt{\frac{1-\ddot{x}}{x+1}}.
\end{eqnarray}
Using the above three integrals, we can derive the general solution of (\ref{exam3}) in the following form
\begin{equation}
x(t)=\frac{I_1-\tilde{I}\sin(I_3+\sqrt{1-I_1}t)} {1-I_1},\label{solmmbb}
\end{equation}
where $\tilde{I}=\sqrt{I_1^2+2I_2-2I_1I_2}$.

\section{Panorama of interconnections}
In the previous section, starting from Lie point symmetries we derived all other quantities. Since the interconnections are global one can consider any other quantity in this family, and derive the rest of them, see the connection diagram (Fig.1). In this section, we consider two such cases, namely (i) integrating factors as starters and (ii) Darboux polynomials as starters and demonstrate the method of deriving all other quantities from them.
\subsection{From integrating factors to others}
In this subsection we demonstrate that starting from the integrating factor we can derive all other quantities for the example (\ref{exam}). Suppose an integrating factor $R_1$ ($R_1=-\frac{1}{x\dot{x}}$) is given as a starter (vide Eq.(\ref{req3})). Then we can get the corresponding null form $U_1$ from Eq.(\ref{met10}) which exactly matches with the first expression given in Eq.(\ref{uuexa}). Substituting $U_1$ in (\ref{ueq}) and rewriting the resultant equation we obtain 
\begin{equation}
D[V_1]-\bigg(\frac{\ddot{x}} {\dot{x}}\bigg)V_1=0.
\end{equation}
A particular solution to this equation is $V_1=-\dot{x}$ which in turn agrees with (\ref{vexa}). Using $V_1$ we can get an expression for $X_1$ through the relation Eq.(\ref{veq}) which also exactly matches with the first expression given in (\ref{xexa}). Once we have $X_1$ and $V_1$ we can obtain the second null form $S_1$ with the help of Eq.(\ref{ueq}). Once we know $S_1$ and $U_1$ we can construct the $\lambda$-symmetry through the expression (\ref{lamus}), that is
\begin{equation}
D[\lambda]+\lambda^2-\frac{\ddot{x}} {\dot{x}}\lambda-\frac{\ddot{x}} {x}=0.
\end{equation}
A particular solution to this equation is $\lambda=\frac{\ddot{x}}{\dot{x}}$ which also agrees with  the first expression given in (\ref{lamex3}). The corresponding $\lambda$-symmetry is $\frac{\partial} {\partial x}$. Using the quantities $R_1$ and $V_1$ in (\ref{rxf}) we can find the Darboux polynomials as given in the first expression in (\ref{dp3v}). In this way we can derive the rest of the quantities from an integrating factor. The exact expressions for the other quantities can be found in Table 1. The associated integral turns out to be $I_1$ as expected. The procedure is exactly the same for the other two integrating factors $R_2$ and $R_3$ which are given in (\ref{req3}). The integrating factors $R_2$ and $R_3$ provide exactly the same expressions $(S_2,U_2,R_2)$ and $(S_3,U_3,R_3)$. We also note here that if we start from some other integrating factor other than the above three then proceeding in the same manner as outlined above one can get their associated null forms and Darboux polynomials. However these expressions do not lead to any new integrals.

\subsection{From Darboux polynomials to others}
Suppose Darboux polynomials which share the same cofactor are given as first information and we have to determine the rest of the quantities. From the Darboux polynomials we can find the integrals using the ratios. We assume that the two Darboux polynomials $F_{12}=x\dot{x}^2$ and $F_3=\dot{x}\ddot{x}$, vide Eqs.(\ref{dp3v}) and (\ref{da4}), are given. Then the ratio gives the integral $I_1=\frac{\ddot{x}}{x\dot{x}}$ (vide Eq.(\ref{int1})). Once we know the integral we can obtain the integrating factor readily by evaluating the last expression given in (\ref{met8}) which exactly matches with the form of $R_1$ given in (\ref{req3}). The method of deriving the rest of the quantities from $R_1$ is outlined in the previous sub-section. In this way, from Darboux polynomials one can derive the rest of the quantities. The procedure is the same for the other Darboux polynomials.

One may assume that the last multipliers are given and then consider the task of determining the other quantities. Using the relation $F=M^{-1}$ we can find the Darboux polynomials. Once Darboux polynomials are known the procedure given in Sec.4(c) may be followed to derive the rest of the quantities. 

\section{Conclusion}
In this work, we have demonstrated the interconnection between Prelle-Singer method (or any one of the methods studied in this paper as the starting point) with the other existing well known methods in the literature such as Jacobi last multipliers, Darboux polynomials, Lie point symmetries, adjoint-symmetries and $\lambda$-symmetries in the case of third-order nonlinear ODEs. For this purpose we started with the PS method. In the PS method, the quantities, namely (i) null forms $U$ and $S$ and (ii) integrating factors $R$ which are determined by three first-order equations play a major role. By introducing suitable transformations to the null forms $U$ and $S$ and to the integrating factor $R$, we have related these three quantities with the above other quantities. While relating the PS method with Darboux polynomials and Jacobi last multiplier, we introduced the transformation $\frac{V}{F}$ in the integrating factor equation. To relate the adjoint-symmetries with the integrating factor we rewrite the three first-order equations in the PS method in terms of a single variable $R$. We then demonstrate that this third-order equation in the variable $R$ coincides with the adjoint-symmetry equation. The difficult and unknown connection between $\lambda$-symmetries and the null forms has been brought out by using the compatibility between the $\lambda$-determining equation and the determining equation for the null forms in the PS method. We have recalled the known connection $\lambda=\frac{D[Q]}{Q}$ to relate the Lie point symmetries and the null forms.  In this way we have established the interconnections between the null forms, integrating factors, adjoint-symmetries, $\lambda$-symmetries, Lie point symmetries, Darboux polynomials and Jacobi last multipliers. We have observed that some of the connections are common for both second-order as well as third-order ODEs, while the others are specifically applicable to third-order ODEs. We have illustrated these interconnections with three definitive examples discussed in the literature. Currently we are extending the above said procedure to $n^{th}$-order nonlinear ODEs. We have obtained some interesting results in this direction. The results will be published in the near future.

\section*{Acknowledgments}
RMS acknowledges the University Grants Commission (UGC-RFSMS), Government of India, for providing a Research Fellowship. The work of MS forms part of a research project sponsored by Department of Science and Technology, Government of India.  The work of ML is supported by a Department of Science and Technology (DST), Government of India, IRHPA research project. ML is also supported by a DAE Raja Ramanna Fellowship and a DST Ramanna Fellowship programme.

\end{document}